# *Effective Medium Model for Graphene Superlattices with Electrostatic and Magnetic Vector Potentials*

David E. Fernandes

*Instituto de Telecomunicações and Department of Electrical Engineering, University of Coimbra, 3030-290 Coimbra, Portugal*

E-mail: dfernandes@co.it.pt

**Abstract**

In this article we develop an effective medium model to characterize the electron wave propagation in graphene based nanostructures with an electrostatic and magnetic vector potentials imposed on their surface. We use a numerical algorithm to determine the effective medium parameters of the heterostructure and calculate the electronic band structure of the system. We apply our formalism to analyze superlattices with solely a magnetic potential and reveal that the response of the structure remains reciprocal and is characterized by a decrease in charge carrier's velocity. We also study the response of superlattices with both potentials superimposed on graphene and show that the response of the system becomes nonreciprocal with a dispersion characterized by a tilted Dirac cone. We demonstrate that it is possible to alternate between a type-I, type-II or type-III Dirac cones by properly tuning the amplitude of the potentials.

# I. Introduction

Graphene is a two-dimensional nanomaterial formed by carbon atoms that are arranged in a honeycomb lattice [1-8]. Over the last decade this material has been on the spotlight of condensed matter physics research due to its remarkable electronic properties. By possessing a relativistic-spectrum, the low-energy electrons in graphene have a linear dispersion and their propagation is determined by a massless Dirac equation [3].

It has been proposed that it is possible to achieve additional control over the propagation characteristics of the electrons in graphene by modifying the original material. These structures are known as graphene superlattices (GSLs) and may be obtained by artificially introducing a new length scale into the system in the form of a periodic potential, either by using an electrostatic potential [9-18] or magnetic vector potential [19-31]. Superlattices characterized by electrostatic potentials may be realized using different techniques, such as with periodically patterned gates, using a crystalline substrate or with the deposition of adatoms on graphene's surface [32-39]. On the other hand, to obtain GSLs with a magnetic vector potential one can use nano-magnetic strips [19-24] or strain-inducing modulations [40]. An electrostatic potential on the surface of graphene can allow for an extreme anisotropic response which can lead to the super-collimation of electron waves [17, 32-34]. Moreover, it can permit extreme wave phenomena such as a perfect lens for matter waves [41, 42] or an electron wormhole [43]. Conversely, a magnetic vector potential can also allow to tailor the electron wave propagation by reducing the charge carriers velocity [24-31] or even providing a way to tilt the energy dispersion of electrons in the medium [44], usually identified as a type-I tilted Dirac cone [45]. Such type of response may be used for valley filtering in *p-n* junctions [46] and to generate photocurrent [47].

The characterization of the propagation of electrons in superlattices with a magnetic vector potential is typically done using a transfer matrix formalism [24-30] which can limit the study to potentials characterized only by constant barriers. Interestingly, in the works proposed in [17, 18, 41-43] the propagation of the electrons in the GSLs with an electrostatic potential was studied under an effective medium formalism, so that granular details of the potential are homogenized [41] and the structure is regarded as a continuum characterized by some effective parameters. Such effective medium techniques can vastly simply the analysis of the problem while simultaneously providing invaluable insight into the physical phenomena taking place in the structure.

The main objective of this work is to develop an effective medium model for superlattices characterized by both an electrostatic potential and a magnetic vector potential. To determine the effective response of the superlattices we use a numerical finite-difference time-domain (FDTD) algorithm based on the numerical tool proposed in Ref. [17]. It is important to mention that FDTD numerical tools such as the ones developed in [17, 18, 48, 49] have been widely used to determine the electron wave propagation in graphene based nanomaterials. To begin with, we apply the numerical algorithm to homogenize a GSL with a magnetic vector potential with a sinusoidal spatial variation and show that, similar to what happens in GSLs with Krönig-Penney type potentials [24-30], the response of the structure is isotropic, with the group velocity of the charge carriers being smaller than in pristine graphene. We demonstrate that for these superlattices the analysis of the problem can be vastly simplified by using an effective Hamiltonian that discards the granular details of the potential and instead considers an effective parameter that is independent of space. This effective parameter may be regarded as an effective Fermi velocity whose value is only dependent on the amplitude of the magnetic vector potential. We also determine the effective response of superlattices with both electrostatic and magnetic vector potentials with sinusoidal spatial variations. Using our effective medium formalism, we demonstrate that the interplay between the magnetic and electric potentials can give rise to an overall nonreciprocal response whose energy dispersion is characterized by a Dirac cone tilted along the direction perpendicular to the stratification of the potentials (type-I Dirac cone). Moreover, we show that for propagation along such direction there is a wide range of combinations of amplitude of the potentials for which the bulk eigenmodes can flow along the same direction and, by properly tuning the amplitude of the potentials, it is even possible to have eigenmodes with a null group velocity. Such dispersion characteristic corresponds to a type-III Dirac cone [45, 50, 51], where one of the bands is flat and the other has a linear dispersion. It has been proposed such dispersion can enhance the superconducting gap in Weyl semimetals [52], and by using the flat band, they can allow for a new platform to study the correlated phases in the structure [53]. Importantly, a type-III Dirac cone marks the transition between type-I and type-II Dirac cones. The type-II dispersion characteristic differs from the type-I from the fact that the Fermi surface is no longer a point, but rather two-crossing lines [54-56]. Such dispersion appears when one of the bands is tilted in such a way that the group velocity of the associated energy eigenmode has the opposite sign that the corresponding value in pristine graphene.

The article is organized as follows. In Sec. II we introduce the homogenization formalism that will be used to characterize the effective medium response of the graphene superlattices. In Sec. III we describe the numerical FDTD algorithm that is used to determine the effective parameters of the superlattice. In Sec. IV the homogenization formalism is applied to characterize the wave propagation in GSLs with solely a magnetic vector potential and in superlattices with both electrostatic and magnetic potentials. Finally, the conclusions are drawn in Sec. V.

## II. Effective Medium Model

In this work we study the electron wave propagation in graphene-based nanomaterials characterized by a periodic electrostatic potential and a periodic magnetic vector potential. Near the $K$ point, the propagation of the charge carriers in graphene superlattices with electrostatic and magnetic vector potentials may be described using the massless Dirac equation:

$$i\hbar \frac{\partial}{\partial t} \mathbf{\psi} = \hat{H} \mathbf{\psi}, \tag{1}$$

where $\hat{H} = v_F \boldsymbol{\sigma} \cdot \left(-i\hbar \nabla - q\mathbf{A}(\mathbf{r})\right) + V(\mathbf{r})$ is the microscopic Hamiltonian operator, $V$ is the microscopic electric potential, $q = -e$ is the electron charge, $\mathbf{A}$ is the magnetic vector potential, $v_F \approx 10^6 \, m/s$ is the Fermi velocity in pristine graphene and $\boldsymbol{\sigma} = \boldsymbol{\sigma}_x \hat{\mathbf{x}} + \boldsymbol{\sigma}_y \hat{\mathbf{y}}$, with $\boldsymbol{\sigma}_x, \boldsymbol{\sigma}_y$ the Pauli matrices. Moreover, $\mathbf{\psi} = \{\Psi_1, \Psi_2\}$ is a two-component pseudospinor, with each component of the pseudospinor associated with a different trigonal sublattice of graphene. In the present work the effects of mass-induced spectral gaps [57-62], as well as strain or spin-orbit coupling effects [63-66], are not taken into account.

Without any loss of generality, we consider that the microscopic potentials have a one-dimensional (1D) spatial variation. Particularly we suppose that $V(\mathbf{r}) = V(x)$ and $\mathbf{A} = A_y(x)\hat{\mathbf{y}}$. The proposed effective model could be readily extended to other (more complex) types of spatial variations.

For the considered spatial variations of the electric and magnetic potentials Eq. (1) may be re-written as:

$$i\hbar \frac{\partial}{\partial t} \psi = v_F \boldsymbol{\sigma} \cdot \left(-i\hbar \nabla + eA_y(x)\hat{\mathbf{y}}\right) \cdot \psi + V(x)\psi. \tag{2}$$

In Ref. [41] it was shown that provided the initial state of the system $\psi(t=0)$ is macroscopic, so that $\langle \psi(t=0) \rangle = \psi(t=0)$, with $\langle \cdots \rangle$ the spatial averaging operator defined as $\langle F \rangle = \frac{1}{A} \int F e^{-i\mathbf{k}\cdot\mathbf{r}} dS$ (where we assume a spatial evolution of the type $e^{i\mathbf{k}\cdot\mathbf{r}}$), then the envelope of the pseudospinor can be accurately calculated from an effective Hamiltonian of the system $\hat{H}_{ef}$, defined as $\hat{H}_{ef} \langle \psi \rangle = \langle \hat{H}\psi \rangle$, which may be written as:

$$\hat{H}_{ef} = \hbar v_F \boldsymbol{\sigma} \cdot \mathbf{k} + e v_F \sigma_y \cdot A_{ef}(\omega, \mathbf{k}) + V_{ef}(\omega, \mathbf{k}). \tag{3}$$

Here we used $i\nabla = \mathbf{k}$, with $\mathbf{k}$ the pseudo-momentum, and introduced the effective magnetic and electric potentials, $A_{ef}$ and $V_{ef}$, which can generally be matrices. Applying the spatial averaging operator to Eq. (2), with $\frac{\partial}{\partial t} = -i\omega$ for a time evolution of the type $e^{-i\omega t}$, we obtain

$$E\langle \psi \rangle = \langle \hat{H}\psi \rangle = \hbar v_F \boldsymbol{\sigma} \cdot \mathbf{k} \langle \psi \rangle + e v_F \sigma_y \cdot \langle A_y \psi \rangle + \langle V\psi \rangle, \tag{4}$$

where $\frac{\partial}{\partial n} = ik_n$, with $n = x, y$ and $E = i\hbar \frac{\partial}{\partial t}$.

To determine the effective Hamiltonian for a fixed energy and pseudo-momentum $\mathbf{k}_0 = (k_{x0}, k_{y0})$ we can calculate the time evolution in the graphene nanomaterial of two linear independent initial electronic states of the form:

$$\psi^{(1)}(\mathbf{r}, t=0) = e^{i\mathbf{k}_0 \cdot \mathbf{r}} \begin{pmatrix} 1 \\ 0 \end{pmatrix}, \tag{5a}$$

$$\psi^{(2)}(\mathbf{r}, t=0) = e^{i\mathbf{k}_0 \cdot \mathbf{r}} \begin{pmatrix} 0 \\ 1 \end{pmatrix}, \tag{5b}$$

and use the Fourier transform and the spatial averaging operator to calculate $\langle \psi^{(n)} \rangle$, $\langle A_y \psi^{(n)} \rangle$ and $\langle V\psi^{(n)} \rangle$ in the frequency and spatial domains for each initial state $\psi^{(n)}$, with $n = 1, 2$. The effective Hamiltonian $\hat{H}_{ef}(\omega, \mathbf{k})$ is then determined by calculating:

$$\hat{H}_{ef}(\omega, \mathbf{k}) = \left[ \langle \hat{H}\psi^{(1)} \rangle ; \langle \hat{H}\psi^{(2)} \rangle \right] \cdot \left[ \langle \psi^{(1)} \rangle ; \langle \psi^{(2)} \rangle \right]^{-1} \quad (6)$$

Similarly, the effective electric and magnetic potentials can also be calculated using:

$$V_{ef}(\omega) = \left[ \langle V\psi^{(1)} \rangle ; \langle V\psi^{(2)} \rangle \right] \cdot \left[ \langle \psi^{(1)} \rangle ; \langle \psi^{(2)} \rangle \right]^{-1} \quad (7)$$

$$A_{ef}(\omega) = \left[ \langle A\psi^{(1)} \rangle ; \langle A\psi^{(2)} \rangle \right] \cdot \left[ \langle \psi^{(1)} \rangle ; \langle \psi^{(2)} \rangle \right]^{-1} \quad (8)$$

It is important to mention that the initial states are chosen in such a way that they are not more localized than the characteristic period of the lattices, such that the pseudo-momentum $\mathbf{k}_0$ associated with these states is within the first Brillouin minizone of the superlattice.

### III. Numerical Algorithm

The calculation of the effective Hamiltonian in the frequency domain requires the calculation of the time evolution of the electronic states in Eq. 5a-b and its Fourier transform. To determine the time-evolution of the initial electronic states we use a FDTD (Finite Differences in the Time Domain) numerical algorithm. The algorithm is based on the numerical tool developed in [17] which was successfully used to study the transport properties of graphene superlattices characterized solely by an electrostatic potential. We begin by separating the Dirac equation (2) for each component of the pseudospinor:

$$\frac{\partial \Psi_1}{\partial t} = -v_F \left( \frac{\partial}{\partial x} - i\frac{\partial}{\partial y} + \frac{eA_y}{\hbar} \right) \Psi_2 + \frac{V}{i\hbar} \Psi_1, \quad (9a)$$

$$\frac{\partial \Psi_2}{\partial t} = -v_F \left( \frac{\partial}{\partial x} + i\frac{\partial}{\partial y} - \frac{eA_y}{\hbar} \right) \Psi_1 + \frac{V}{i\hbar} \Psi_2. \qquad (9b)$$

To obtain the time update equations in an explicit form we discretize the spatial domain into a rectangular grid, such that the consecutive nodes along the *x*- and *y*-directions are separated by a distance $\Delta_x$ and $\Delta_y$, as depicted in Fig. 1a. Furthermore, each component of the pseudospinor is sampled at instants of time separated by the time step $\Delta_t$. This allows us to write the pseudospinor components as $\Psi(x,y,t) = \Psi(p\Delta_x, q\Delta_y, n\Delta_t) \equiv \Psi(p,q,n)$ and permits using a finite differences method to calculate the partial derivatives in Eq(9) such that:

$$\partial_l \Psi(i) = \frac{\Psi\left(i+\frac{1}{2}\right) - \Psi\left(i-\frac{1}{2}\right)}{\Delta_l}, \qquad (10)$$

with $l = x, y, t$.

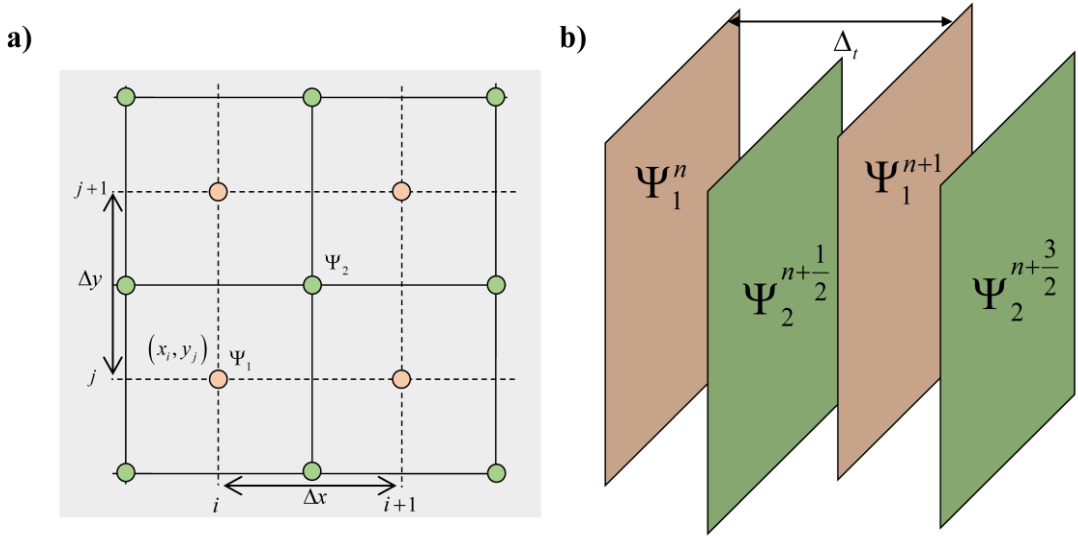

**Fig. 1** (color online) **a)** The graphene superlattice spatial domain is discretized into a rectangular grid with a finite number of nodes spaced by $\Delta_x$ along the *x*-direction and $\Delta_y$ along the *y*-direction. The pseudospinor components $\Psi_1$ and $\Psi_2$ are defined on staggered subgrids so that $\Psi_1$ is defined over the nodes $(p,q)$ and $\Psi_2$ is defined at the nodes $(p+1/2, q+1/2)$, shifted by a half-grid period. **b)** The time domain is sampled at time intervals separated by a time step $\Delta_t$. Similarly to the spatial domain discretization scheme, the pseudospinor components $\Psi_1$ and $\Psi_2$ are defined on staggered subgrids shifted by $\Delta_t/2$.

We also consider that the two components of the pseudospinor are defined on staggered grids in space and time so that $\Psi_1(p,q,n)$ and $\Psi_2\left(p+\frac{1}{2},q+\frac{1}{2},n+\frac{1}{2}\right)$, as shown in Fig. 1a-b. Applying these principles to Eqs. 9a-b leads to the following update equations:

$$\Psi_{1,p,q}^{n+1}\left(1-\frac{V_{p,q}}{i\hbar 2}\Delta_t\right) = \Psi_{1,p,q}^{n}\left(1+\frac{V_{p,q}}{i\hbar 2}\Delta_t\right) - v_F \Delta_t \left[ +\frac{eA_{y_{p,q}}}{\hbar} \Psi_{2,p,q}^{n+\frac{1}{2}} + \left(\frac{1}{2\Delta_x} - i\frac{1}{2\Delta_y}\right)\Psi_{2,p+\frac{1}{2},q+\frac{1}{2}}^{n+\frac{1}{2}} \right.$$

$$\left. -\left(\frac{1}{2\Delta_x} + i\frac{1}{2\Delta_y}\right)\Psi_{2,p-\frac{1}{2},q+\frac{1}{2}}^{n+\frac{1}{2}} + \left(\frac{1}{2\Delta_x} + i\frac{1}{2\Delta_y}\right)\Psi_{2,p+\frac{1}{2},q-\frac{1}{2}}^{n+\frac{1}{2}} - \left(\frac{1}{2\Delta_x} - i\frac{1}{2\Delta_y}\right)\Psi_{2,p-\frac{1}{2},q-\frac{1}{2}}^{n+\frac{1}{2}} \right]$$

(11a)

$$\Psi_{2,p+\frac{1}{2},q+\frac{1}{2}}^{n+\frac{1}{2}}\left(1-\frac{V_{p+\frac{1}{2},q+\frac{1}{2}}}{i\hbar 2}\Delta_t\right) = \Psi_{2,p+\frac{1}{2},q+\frac{1}{2}}^{n-1/2}\left(1+\frac{V_{p+\frac{1}{2},q+\frac{1}{2}}}{i\hbar 2}\Delta_t\right) - v_F \Delta_t \left[ -\frac{eA_{y_{p,q}}}{\hbar}\Psi_{1,p+\frac{1}{2},q+\frac{1}{2}}^{n} + \left(\frac{1}{2\Delta_x} + i\frac{1}{2\Delta_y}\right)\Psi_{1,p+1,q+1}^{n} \right.$$

$$\left. -\left(\frac{1}{2\Delta_x} - i\frac{1}{2\Delta_y}\right)\Psi_{1,p,q+1}^{n} + \left(\frac{1}{2\Delta_x} - i\frac{1}{2\Delta_y}\right)\Psi_{1,p+1,q}^{n} - \left(\frac{1}{2\Delta_x} + i\frac{1}{2\Delta_y}\right)\Psi_{1,p,q}^{n} \right]$$

(11b)

With $V_{p,q} = V(p\Delta_x, q\Delta_y)$ and $A_{y_{p,q}} = A_y(p\Delta_x, q\Delta_y)$. Importantly, this discretization scheme of the update equations requires the value of the pseudospinor components in subgrid points where they are not defined. In particular, the value value $\Psi_2(p,q)$ is necessary in Eq. (11a), while the update equation 11b requires the value of $\Psi_1\left(p+\frac{1}{2},q+\frac{1}{2}\right)$. To obtain such values we assume that the wavefunction varies slowly in space so that the pseudospinor component values in points of space that lie outside the grid nodes can be determined by the average of its neighboring nodes. In that case one can use:

$$\Psi_2(p,q) \approx \frac{1}{4}\left(\Psi_2\left(p+\frac{1}{2},q-\frac{1}{2}\right)+\Psi_2\left(p-\frac{1}{2},q+\frac{1}{2}\right)+\Psi_2\left(p+\frac{1}{2},q+\frac{1}{2}\right)+\Psi_2\left(p-\frac{1}{2},q-\frac{1}{2}\right)\right), \quad (12a)$$

$$\text{and } \Psi_1\left(p+\frac{1}{2},q+\frac{1}{2}\right) \approx \frac{1}{4}\left(\Psi_1(p+1,q)+\Psi_1(p,q+1)+\Psi_1(p+1,q+1)+\Psi_1(p,q)\right). \quad (11b)$$

To determine the effective Hamiltonian it necessary to calculate the Fourier transform of $\langle\psi(t)\rangle$, $\langle A_y\psi(t)\rangle$ and $\langle V\psi(t)\rangle$. As shown in other works dealing with time-domain homogenization techniques of artificial structured media [67], to ensure the convergence of the Fourier transform for a given frequency $\omega$, at each time iteration $n=0,...,N$ the quantities $\langle\psi(n\Delta t)\rangle$, $\langle A_y\psi(n\Delta t)\rangle$ and $\langle V\psi(n\Delta t)\rangle$ must be multiplied by a time decaying exponential of the form $e^{\omega''n\Delta t}$, with $\omega=\omega'+i\omega''$, so that the term $\omega''$ represents some small losses in the system. Importantly, the total number of iterations $N$ must be sufficiently high so that $e^{\omega''N\Delta t}\approx 0$. Typically, the total number of iterations is on the order of $N\approx 2\pi/\omega''\Delta_t$ [67].

Without any loss of generality in all calculations performed in this work we consider that both potentials have the same spatial period $a$ and that the distance between adjacent nodes is the same, i.e. $\Delta_x=\Delta_y=\Delta$. Moreover, we determine the time evolution of the initial states in a region of space containing one period of the potentials and apply Bloch boundary condition at the edge of the computational domain. In Appendix A we analyze the stability conditions of the proposed FDTD algorithm and show that the maximum value of the time-step depends both on the distance between adjacent spatial nodes and the maximum amplitude of the magnetic potential.

## IV. Numerical results

In what follows, we use our homogenization algorithm to determine the band diagram and effective medium parameters of graphene superlattices with electrostatic and magnetic vector potentials.

It was shown in [41] that provided the microscopic Hamiltonian of the graphene nanomaterial does not vary in time, the electronic band structure of the material close to the $K$ point can be computed directly from the effective Hamiltonian. This property is a consequence of the energy eigenstates of the system calculated using $\hat{H}_{ef}$ being equal to the eigenstates of the microscopic Hamiltonian [41].

In this work we consider an electrostatic potential with a sinusoidal-type spatial variation of the form $V(x) = V_{osc} \sin\left(\frac{2\pi x}{a}\right)$ and a periodic magnetic induction field given by $\mathbf{B} = B_z(x)\hat{\mathbf{z}} = \frac{2\pi}{a_1} A_{osc} \cos\left(\frac{2\pi x}{a_1}\right)\hat{\mathbf{z}}$, so that it varies along the $x$-direction but is oriented along the $z$-direction (perpendicular to the propagation plane). The magnetic induction field it is on average null, i.e. $\frac{1}{a_1}\int_0^{a_1} B_z(x)\,dx = 0$. Since the magnetic field is related to the magnetic vector potential as $\mathbf{B} = \nabla \times \mathbf{A}$, it follows that $\mathbf{A} = A_y(x)\hat{\mathbf{y}}$, with $A_y(x) = A_{osc} \sin\left(\frac{2\pi x}{a}\right)$, which is also periodic with zero spatial average.

Similar to [41], to calculate the band diagram of the structures we expand the effective potentials $A_{ef}$, $V_{ef}$ as a Taylor series of the first order, such that:

$$A_{ef}(\omega, \mathbf{k}) = A_{ef}(\omega, 0) + \frac{\partial A_{ef}(\omega, 0)}{\partial k_x} k_x + \frac{\partial A_{ef}(\omega, 0)}{\partial k_y} k_y \text{ and} \tag{13}$$

$$V_{ef}(\omega, \mathbf{k}) = V_{ef}(\omega, 0) + \frac{\partial V_{ef}(\omega, 0)}{\partial k_x} k_x + \frac{\partial V_{ef}(\omega, 0)}{\partial k_y} k_y, \tag{14}$$

and use our numerical algorithm to calculate the effective Hamiltonian (given by Eq. 4). Note that in general $A_{ef}(\omega,\mathbf{k})$ and $V_{ef}(\omega,\mathbf{k})$ are not scalars.

From hereon we consider that the spatial grid is discretized using a node spacing $\Delta = a/50$ and the time step is $\Delta_t = 0.3\Delta/v_F$. Since the effective response of graphene superlattices characterized solely by electrostatic potentials was already thoroughly discussed in [17, 18, 41], we restrict our attention to superlattices with only magnetic vector potential and structures with both magnetic and electrostatic potentials.

### A) Superlattices with Magnetic Potential

We begin by calculating the response of superlattices characterized solely by a magnetic vector potential. Particularly, we determine the effective potential $A_{ef}(\omega,\mathbf{k})$ of the nanomaterial for some amplitudes of the magnetic potential. For this GSL it is immediate that the effective electric potential is null, i.e. $V_{ef}(\omega,\mathbf{k}) = 0$.

Our numerical results showed that to an excellent approximation $A_{ef}(\omega,0) \approx \xi \boldsymbol{\sigma}_y$, with $\xi$ a real value depicted in Fig. 2a. Interestingly, it is seen that for low-energy excitations this value varies linearly with the energy on the carriers. Moreover, we also verified that $\frac{\partial A_{ef}(\omega,0)}{\partial k_x} \approx i\beta \boldsymbol{\sigma}_z$, with $\boldsymbol{\sigma}_z$ the Pauli matrix, and that $\frac{\partial A_{ef}(\omega,0)}{\partial k_y} \approx \chi \mathbf{1}$.

Here $\beta \approx \chi$ are real constants almost independent of the carriers energy for low-energy excitations, as shown in Fig. 2b-c, respectively, for some representative amplitudes of the potential $A_{osc}$.

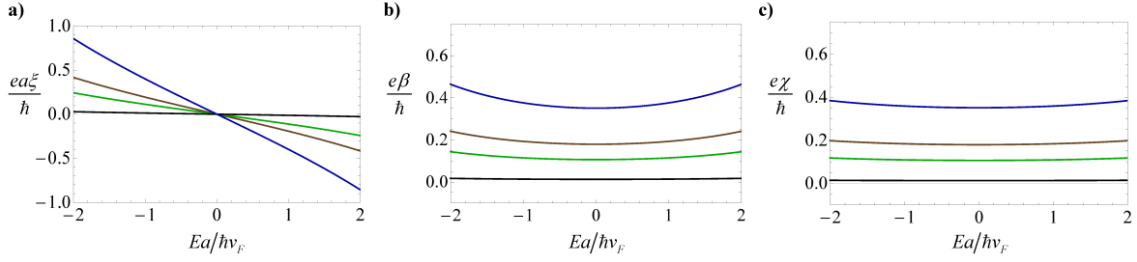

**Fig. 2** Normalized effective parameters of the graphene superlattice as a function of the normalized energy: **a)** $ea\xi/\hbar$, **b)** $e\beta/\hbar$ and **c)** $e\chi/\hbar$, for a magnetic vector potential with amplitude $eA_{osc}a/\hbar = 1.0$ (black line), $eA_{osc}a/\hbar = 3.0$ (dark green line), $eA_{osc}a/\hbar = 4.0$ (brown line) and $eA_{osc}a/\hbar = 6.0$ (blue line).

Hence, for low energy excitations, the effective magnetic potential may be approximated by:

$$A_{ef}(\omega,\mathbf{k}) \approx \xi(\omega)\boldsymbol{\sigma}_y + i\beta k_x \boldsymbol{\sigma}_z + \chi k_y \mathbf{1} \quad (15)$$

Inserting this expression in Eq. 3, allows us to write the effective Hamiltonian as:

$$\hat{H}_{ef} = \upsilon_0 E \mathbf{1} + \upsilon \hbar v_F \boldsymbol{\sigma} \cdot \mathbf{k}, \quad (16)$$

with $\upsilon_0 E = ev_F \xi$ and $\upsilon = 1 - \beta e/\hbar \approx 1 - \chi e/\hbar$. The energy dispersion of the superlattice may then be calculated from the eigenvalue problem:

$$E\Psi = \hat{H}_{ef} \cdot \Psi, \quad (17)$$

and it is given by $|E - \upsilon_0 E| = \hbar v_F |\upsilon| \sqrt{k_x^2 + k_y^2}$. Since $\upsilon_0$ and $\upsilon$ are weakly dependent on the energy, the previous expression can be further simplified into

$$|E| = \hbar v_{F,eff} |\mathbf{k}|, \quad (18)$$

by defining an effective Fermi velocity $v_{F,eff} = v_F \dfrac{|\upsilon|}{|1-\upsilon_0|}$. Considering that $\upsilon_0$ is a negative value (proportional to the slope of the curves in Fig. 2a) and that $\upsilon$ is smaller than unity because $\beta, \chi > 0$, it is expected that $v_{F,eff}$ can be significantly smaller than the Fermi velocity. To verify the accuracy of our effective medium model we overlapped in Fig. 3a-b the "exact" band diagram for propagation along the *x*-direction ($k_y = 0$) and the *y*-direction ($k_x = 0$), with the corresponding results calculated using our simplified effective formalism. The band diagram was calculated for a graphene superlattice characterized by a magnetic potential with amplitude $eA_{osc}a/\hbar = 4.0$. As seen, for low energy excitations both results have nearly exact agreement.

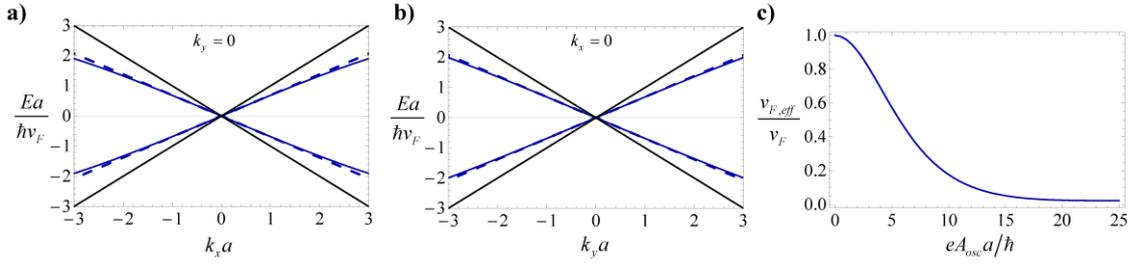

**Fig. 3 a)** and **b)** Dispersion of the energy eigenstates of a graphene superlattice with a magnetic vector potential $eA_{osc}a/\hbar = 4.0$ for propagation along the *x* and *y* directions, respectively. The blue dashed curves represent the "exact" energy dispersion of the GSL, the blue solid curves correspond to the dispersion of the GSL calculated using the effective parameter $v_{F,eff}$ and the black solid curves show the dispersion of pristine graphene. **c)** Normalized effective Fermi velocity $v_{F,eff}/v_F$ as a function of the normalized magnetic vector potential amplitude $eA_{osc}a/\hbar$.

The results shown in Fig. 3a-b also show that even in the presence of a magnetic potential, which the breaks time-reversal symmetry of the structure, the response of these graphene superlattices remains isotropic and reciprocal for low energy excitations, and without a bandgap. Moreover, we see that the group velocity of the carriers in the superlattice is exactly equal to the effective Fermi velocity $v_{g,x} = v_{g,y} = \frac{1}{\hbar}\frac{dE}{dk} = v_{F,eff}$.

Indeed, by comparing these results with the band diagram of pristine graphene, depicted as black curves in Fig. 3a-b, it is confirmed that the effective Fermi velocity is smaller than the Fermi velocity in pristine graphene. In in Fig. 3c we show the effect of the amplitude of the magnetic potential on the effective Fermi velocity. The results reveal that the carrier's velocity can be severely reduced from the Fermi velocity as the amplitude of the magnetic vector potential increases. Hence, by precisely tailoring the magnetic field distribution we can control the charge velocity in the medium. These results go in line with previous studies [24-30], which using a transfer matrix formalism, demonstrated the effect of magnetic potential barriers in the carrier velocity properties. Moreover, the decrease in the effective Fermi velocity can be regarded as the magnetic equivalent to the Klein tunneling effect in graphene superlattices with a periodic electrostatic potential, which is originated from pseudospin nature of the eigenstates in the GSL [31].

### B) Superlattices with Magnetic and Electric Potentials

The analysis we did in the previous section revealed that imposing a 1D magnetic potential with zero-spatial average on the surface of graphene leads to a reciprocal and isotropic response, wherein the charge carriers group velocity is decreased as the amplitude of the magnetic potential increases. On the other hand, it was demonstrated in [17, 41] that in graphene superlattices with 1D electrostatic potential with zero spatial average, the transport properties of the electrons can also be modified so that they have a preferred direction of propagation, i.e. the effective medium behaves as an anisotropic medium. In what follows, we use our homogenization model to study the response of superlattices characterized by both a 1D magnetic and a 1D electric potential with zero spatial average.

As in the previous section, we start by calculating the effective potentials of the GSL given by Eqs. (13)-(14). As a leading example, we consider a superlattice characterized by a magnetic vector potential with amplitude $eA_{osc}a/\hbar = 4$ and an electrostatic potential with amplitude $V_{osc}a/\hbar v_F = 5$. Our numerical results show to an excellent approximation that the Taylor expansion of the effective potentials may be written as:

$$A_{ef}(\omega,\mathbf{k}) = \alpha_0 \mathbf{1} + \beta_0 \boldsymbol{\sigma}_y + \boldsymbol{\sigma}_x \cdot (\alpha_1 \mathbf{1} + \beta_1 \boldsymbol{\sigma}_y) k_x + (\alpha_2 \mathbf{1} + \beta_2 \boldsymbol{\sigma}_y) k_y \qquad (19)$$

$$V_{ef}(\omega,\mathbf{k}) = \chi_0 \mathbf{1} + \delta_0 \boldsymbol{\sigma}_y + \boldsymbol{\sigma}_x \cdot (\chi_1 \mathbf{1} + \delta_1 \boldsymbol{\sigma}_y) k_x + (\chi_2 \mathbf{1} + \delta_2 \boldsymbol{\sigma}_y) k_y, \qquad (20)$$

With $\alpha_i, \beta_i, \chi_i, \delta_i$, with $i = 0,1,2$, real-valued scalars whose energy dependence is shown in Fig. 4.

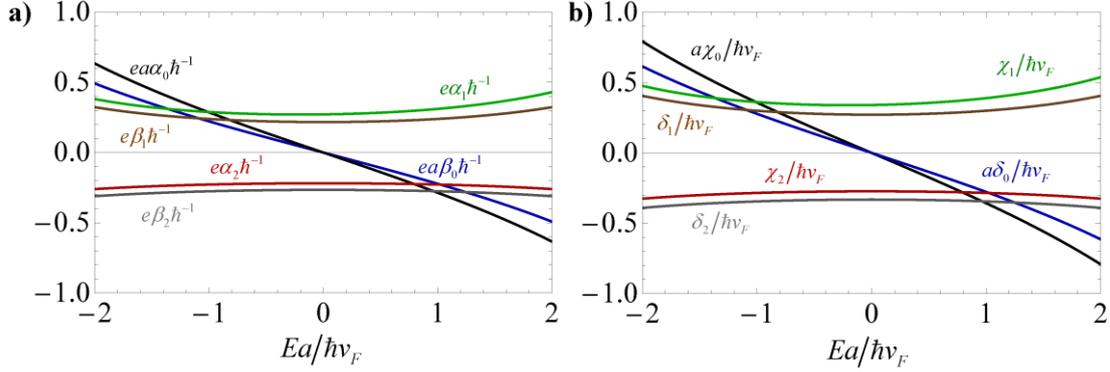

**Fig. 4 a)** Normalized effective magnetic potential parameters $ea\alpha_0 \hbar^{-1}$, $ea\beta_0 \hbar^{-1}$, $e\alpha_1 \hbar^{-1}$, $e\beta_1 \hbar^{-1}$, $e\alpha_2 \hbar^{-1}$, $e\beta_2 \hbar^{-1}$ as a function of the normalized energy for a graphene superlattice for a characterized by a magnetic vector potential with amplitude $eA_{osc} a/\hbar = 4$ and an electrostatic potential with amplitude $V_{osc} a/\hbar v_F = 5$. **b)** Similar to **a)** but for the normalized effective electrostatic potential parameters $ea\chi_0 \hbar^{-1}$, $ea\delta_0 \hbar^{-1}$, $e\chi_1 \hbar^{-1}$, $e\delta_1 \hbar^{-1}$, $e\chi_2 \hbar^{-1}$, $e\delta_2 \hbar^{-1}$.

The results in Fig. 4 show that the zeroth order coefficients of the Taylor expansion of the potentials $\alpha_0, \beta_0, \chi_0, \delta_0$ vary linearly with the energy of the electrons. On the other hand, the first order terms $\alpha_i, \beta_i, \chi_i, \delta_i$, with $i = 1,2$, are almost independent of the energy. Interestingly, our simulation results also show that the effective potentials are linked to each other through the following relations:

$$\frac{ev_F \alpha_0}{E} \approx \frac{\delta_0}{E} \approx -\frac{e\alpha_1}{\hbar} \approx \frac{e\beta_2}{\hbar} \approx \frac{\chi_2}{\hbar v_F} \approx -\frac{\delta_1}{\hbar v_F} = c_0, \qquad (21a)$$

$$\frac{ev_F \beta_0}{E} \approx -\frac{e\beta_1}{\hbar} \approx \frac{e\alpha_2}{\hbar} = c_1, \qquad (21b)$$

$$\frac{\chi_0}{E} \approx -\frac{\chi_1}{\hbar v_F} \approx \frac{\delta_2}{\hbar v_F} = c_2. \qquad (21c)$$

Surprisingly, these relations show that it is possible to describe the spatially dispersive response of the potentials at the expense of the non-spatially dispersive terms. For this example, it is found that $c_0 \approx -0.277$, $c_1 \approx -0.215$ and $c_2 \approx -0.269$. Using the effective potentials (19)-(20) together with Eqs. 21a-c, we can write the effective Hamiltonian (Eq. 3) as:

$$\hat{H}_{ef} = \hbar v_F \boldsymbol{\sigma} \cdot \mathbf{k} + 2c_0 E \boldsymbol{\sigma}_y + (c_1 + c_2) E \mathbf{1} + (c_1 - c_2) \hbar v_F k_x \boldsymbol{\sigma}_x + k_y \hbar v_F \left( (c_1 + c_2) \boldsymbol{\sigma}_y + 2c_0 \mathbf{1} \right). \quad (22)$$

The energy dispersion of the modes supported in the superlattice is obtained from the eigenvalue problem in Eq. (17) using the effective Hamiltonian given by Eq. (22). The corresponding band diagram is shown in Fig. 5a and reveals that the response of the superlattice is no longer reciprocal, with the band diagram being tilted along the $k_y$.

To have a better understanding of the structure's nonreciprocal response effect in the propagation of the electrons we focus our attention on propagation along the $x$- and $y$-directions. Let us begin by studying the propagation along $x$ ($k_y = 0, k_x \neq 0$). In that case, the energy dispersion of the modes is simply given by:

$$|E| = \frac{\hbar v_F (1 + c_1 - c_2)|k_x|}{\sqrt{\left|4c_0^2 - (1 - c_1 - c_2)^2\right|}} \qquad (23)$$

The corresponding band diagram is show in Fig. 5b, where we also depict the "exact" band diagram.

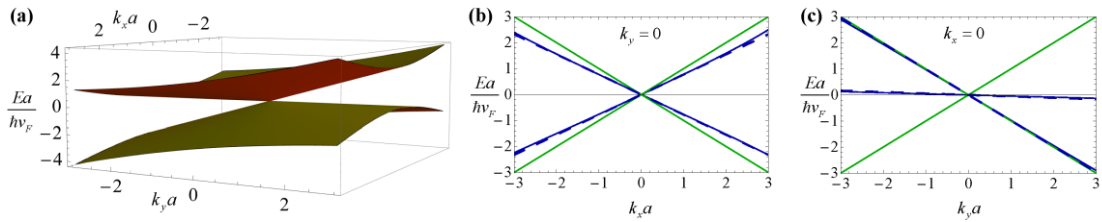

**Fig. 5 a**) Exact energy dispersion of the considered graphene superlattice **b**) and **c**) Dispersion of the energy eigenstates for propagation along the $x$ and $y$ directions, respectively. The blue dashed curves represent the "exact" energy dispersion of the GSL, the blue solid curves correspond to the dispersion of the GSL calculated using the simplified effective medium formalism and the green solid curves show the dispersion of pristine graphene. In all panels the GSL is characterized by a magnetic vector potential with amplitude $eA_{osc}a/\hbar = 4.0$ and an electrostatic potential with amplitude $V_{osc}a/\hbar v_F = 5.0$.

Our effective medium model results have a very good agreement with the exact band diagram showing that for propagation along the *x*-direction the results are comparable to the band diagram obtained for superlattices with solely a magnetic potential (see Fig. 3a). Indeed, for propagation along the *x*-direction, the response of the structure is reciprocal and characterized by a group velocity smaller to that of pristine graphene (whose response is depicted as green curves in Fig. 5a), so that $v_{g,x} = \hbar^{-1} \partial E/\partial k_x \approx 0.77 v_F$.

To determine the band diagram for propagation perpendicular to the direction of the stratification of the potentials, i.e. for propagation along the *y*-direction (with $k_x = 0$, $k_y \neq 0$), we follow a similar procedure and calculate the eigenvalue problem in Eq. (17) considering the effective Hamiltonian given by Eq. (22) with $k_x = 0$. The problem yields two solutions (eigenmodes), whose energy dispersion is given by:

$$E^{(1)} = \hbar v_F k_y \frac{2c_0 - c_1 - c_2 - 1}{2c_0 - c_1 - c_2 + 1} \tag{24a}$$

$$E^{(2)} = \hbar v_F k_y \frac{2c_0 + c_1 + c_2 + 1}{-2c_0 - c_1 - c_2 + 1} \tag{24b}$$

Clearly, for a fixed pseudo-momentum $k_y$ both solutions are not symmetric. In Fig. 5c we represent the band diagram calculated using the effective medium formalism and overlap the results with both the exact band diagram of the superlattice and the band diagram of pristine graphene. As seen, both results predict that the superlattice response is vastly different from that of pristine graphene. The response of the structure is nonreciprocal as for a fixed energy both bulk modes are characterized by wave vectors $k_y$ that have the same sign. Additionally, the band diagram reveals that it is possible to have unidirectional bulk modes as both bands have negative (but distinct) group velocity, i.e. $v_{g,y}^{(1)} = \hbar^{-1} \partial E^{(1)}/\partial k_y \neq v_{g,y}^{(2)} = \hbar^{-1} \partial E^{(2)}/\partial k_y$, consistent with a type-II Dirac cone dispersion characteristic. While one of the bands ($E^{(1)}(k_y)$) follows closely the original band of the pristine graphene, the other one ($E^{(2)}(k_y)$) is tilted towards the origin $E = 0$. Indeed, our results suggest that it may be possible obtain an eigenmode characterized by a flat band so that the group velocity is precisely equal to zero $v_{g,y} = 0$, corresponding to a static wave. Importantly, this type of dispersion characteristic is usually identified as a type-III Dirac cone [45] where the energy dispersion consists of one flat band while the other has a liner dispersion [45, 50, 51]. Clearly, this is an effect of the interplay between the electrostatic and magnetic vector potentials in the dynamics of the charge carriers in the superlattice. In Fig. 6a we show the group velocities $v_{g,y}^{(1)}, v_{g,y}^{(2)}, v_{g,x}$ as a function of the amplitude of the magnetic potential for the fixed amplitude of the electric potential $V_{osc} a/\hbar v_F = 5$. It is seen that increasing the amplitude of the magnetic potential decreases the group velocity along the *x*-direction, similar to the previous studied superlattice with $V_{osc} = 0$. On the other hand, the effects of changing $A_{osc}$ in the group velocities $v_{g,y}^{(1)}, v_{g,y}^{(2)}$ are far more pronounced. To begin with, we note that when $A_{osc} = 0$ the response is that of a superlattice with electric potential, which is characterized by an anisotropic response [17, 41], so that $v_{g,y}$ can significantly smaller than the Fermi velocity. Moreover, we see the group velocity $v_{g,y}^{(1)}$ decreases as the amplitude of the magnetic potential increases, reaching a minimum for $eA_{osc} a/\hbar \approx 4.71$, at which point it is equal to the Fermi velocity, and then it starts increasing. In contrast, for the other eigenmode, its group velocity $v_{g,y}^{(2)}$ decreases as $A_{osc}$ increases, reaching a null value for a magnetic potential with amplitude $eA_{osc}a/\hbar \approx 3.36$. Such combination of amplitudes leads to dispersion characterized by a type-III Dirac cone, which crucially, marks the transition point where the dispersion changes from a type-I Dirac cone (for $eA_{osc}a/\hbar < 3.36$) into a type-II (when $eA_{osc}a/\hbar > 3.36$), where both eigenmodes flow along the same direction.

To have a complete understanding of the interplay between both potentials in the carriers velocity in the superlattice, we numerically calculated the group velocities $v_{g,x}, v_{g,y}^{(1)}, v_{g,y}^{(2)}$ while simultaneously varying $A_{osc}$ and $V_{osc}$. The results are shown in Fig. 6b-d, respectively.

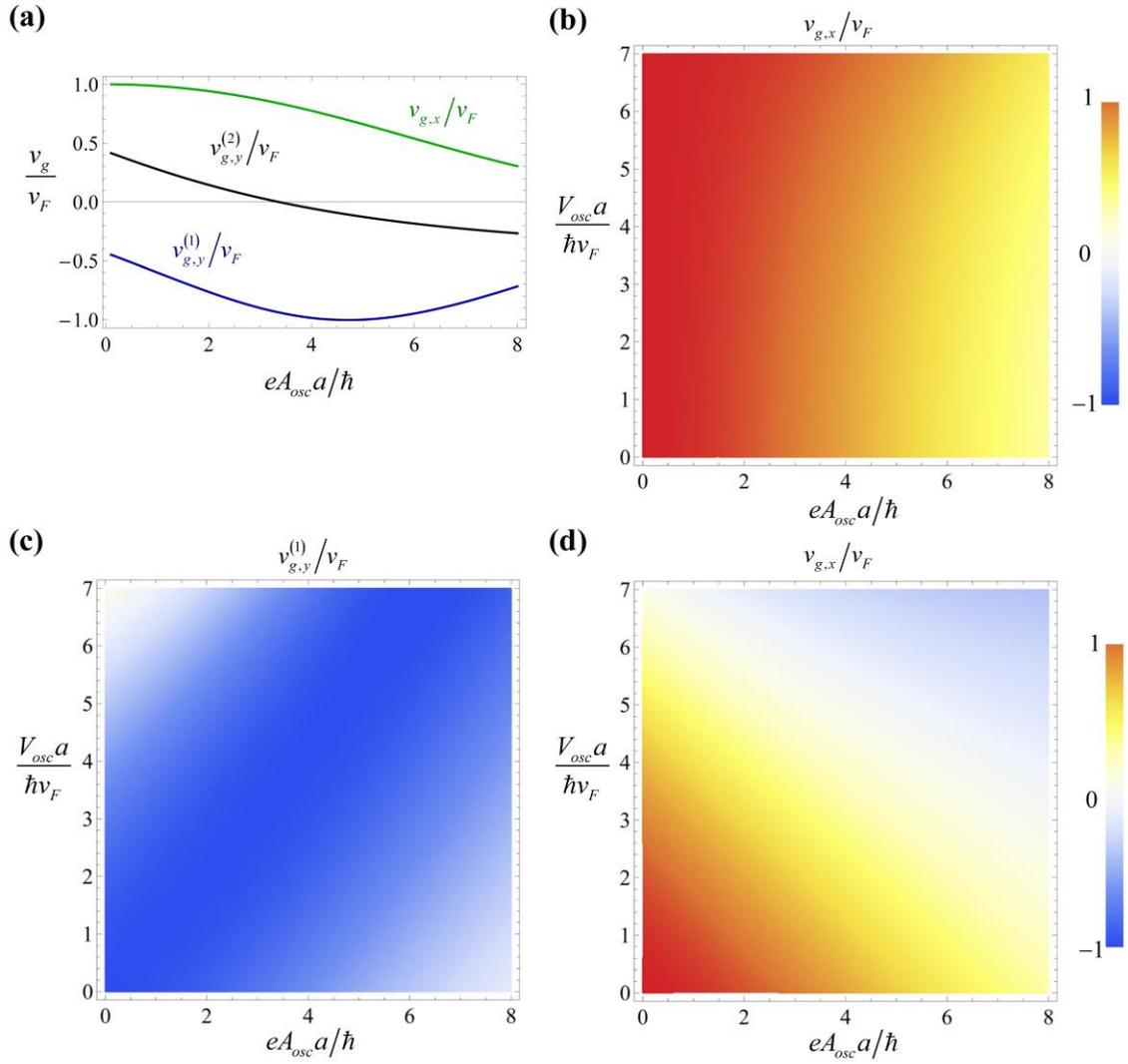

**Fig. 6 a)** Normalized group velocities along the $x$-direction $v_{g,x}/v_F$ and along the $y$-direction $v_{g,y}^{(1)}/v_F$ and $v_{g,y}^{(2)}/v_F$ of the bulk eigenmodes of a GSL is characterized by an electrostatic potential with amplitude $V_{osc}a/\hbar v_F = 5.0$ as a function of the normalized amplitude of the magnetic vector potential $eA_{osc}a/\hbar$. **b)** Normalized group velocities along the $x$-direction $v_{g,x}/v_F$ of the bulk eigenmodes as a function of the normalized amplitudes of the magnetic vector potential $eA_{osc}a/\hbar$ and electrostatic potential $V_{osc}a/\hbar v_F$. **c)** and **d)** Similar to **b)** but for the normalized group velocities of the bulk eigenmodes that propagate along the $y$-direction $v_{g,y}^{(1)}/v_F$ and $v_{g,y}^{(2)}/v_F$, respectively.

The results depicted in Fig. 6b reveal that $v_{g,x}$ is unaffected by a variation in $V_{osc}$ when $A_{osc} = 0$ due to the Klein tunneling effect, and that when $A_{osc} \neq 0$ the carriers' velocity is decreased. Interestingly, our results suggest that for a fixed magnetic potential $v_{g,x}$ tends to increase as the amplitude of the electric potential increases.

For propagation along the *y*-direction, the response of the structure to variations in both potentials is more complex. In Fig. 6c it is seen that the group velocity $v_{g,y}^{(1)}$ follows closely the response of pristine graphene when both potentials are similarly valued (using normalized quantities) but decreases significantly when one of the potentials is significantly stronger than the other. On the other hand, the results associated with the propagation properties of the other eigenmode, shown in Fig. 6d, show that $v_{g,y}^{(2)}$ is progressively reduced when both $A_{osc}$ and $V_{osc}$ increase. Hence, there is a vast combination of potentials that can result in static waves ($v_{g,y}^{(2)} = 0$), i.e. a dispersion characteristic consistent with a type-III Dirac cone, and also bulk unidirectional eigenmodes ($v_{g,y}^{(1)}, v_{g,y}^{(2)} < 0$) (type-II Dirac cone), so that by properly tuning the potentials we are able to precisely control the direction of propagation of the carriers in the superlattice.

## V. Conclusions

We developed a homogenization model to determine the effective response of superlattices with magnetic vector potential and electrostatic potential. We used a numerical FDTD algorithm to apply this formalism and study the propagation properties of the charge carriers in graphene superlattices characterized by 1D magnetic and electric potentials with a sinusoidal-type spatial variation. We demonstrated that when the GSL has only a magnetic vector potential, the effective Hamiltonian of the structure can be drastically simplified by neglecting the granular details of the potential and considering solely an effective Fermi velocity that is smaller to that of pristine graphene. We also demonstrated that when both potentials are present in the superlattice the response of the structure becomes nonreciprocal and is characterized by a dispersion characteristic consisting of a tilted Dirac cone. Particularly, we showed that for propagation perpendicular to the stratification of the potentials the GSL supports two eigenmodes whose energy dispersion, for a fixed pseudo-momentum, is not linked by an odd symmetry. We showed that in such materials we can obtain extreme wave phenomena such as energy flat bands and regimes where both eigenmodes flow along the same direction. We envision that by properly tuning the potentials the proposed GSL structure can be operated in regimes characterized by type-I, type-II or type-III Dirac cones.

## Appendix A: Stability of the FDTD Algorithm

In a FDTD numerical algorithm the calculations remain stable provided the time step is small enough, below a give threshold [68]. In what follows we address the stability of the proposed numerical FDTD algorithm to determine the time evolution of the waves in the graphene superlattices. For simplicity we assume that the medium is spatially homogeneous ($V$ and $A_y$ are independent of the spatial coordinates) in the update equations 11a-b. Our aim is to characterize the stationary states of the system. Thus, we look for plane-wave type solutions of Eq. (11) with:

$$\begin{pmatrix} \Psi_{1,p+1,q}^{n} \\ \Psi_{2,p+1/2,q+1/2}^{n+1/2} \end{pmatrix} = \xi_p \begin{pmatrix} \Psi_{1,p,q}^{n} \\ \Psi_{2,p-1/2,q+1/2}^{n+1/2} \end{pmatrix}, \quad \text{(A1a)}$$

$$\begin{pmatrix} \Psi_{1,p,q+1}^{n} \\ \Psi_{2,p+1/2,q+1/2}^{n+1/2} \end{pmatrix} = \xi_q \begin{pmatrix} \Psi_{1,p,q}^{n} \\ \Psi_{2,p+1/2,q-1/2}^{n+1/2} \end{pmatrix}, \quad \text{(A1b)}$$

where $\xi_p = e^{i\theta_p}$ and $\xi_q = e^{i\theta_q}$ are the spatial phase-shifts between consecutive nodes. Furthermore, we consider a time variation of the type $\Psi_{1,p,q}^{n+1} = \lambda \Psi_{1,p,q}^{n}$ and

$\Psi_{2,p,q}^{n+1/2} = \lambda \Psi_{2,p,q}^{n-1/2}$ where $\lambda$ is a function of the spatial phase-shifts $(\xi_p, \xi_q)$. Hence, the proposed FDTD algorithm is stable as long as $|\lambda| \leq 1$ for arbitrary values of $\xi_p$ and $\xi_q$ with $|\xi_p| = |\xi_q| = 1$. Inserting Eq. A1a-b into Eq. 11a-b and using simple mathematical manipulations we obtain the following system written in a matrix form:

$$\begin{pmatrix} \frac{1}{\lambda}\left(\lambda - 1 - \frac{V}{i\hbar 2}\Delta_t(\lambda+1)\right) & v_F \Delta_t D_- \\ v_F \Delta_t D_+ & \left(\lambda - 1 - \frac{V}{i\hbar 2}\Delta_t(\lambda+1)\right) \end{pmatrix} \begin{pmatrix} \Psi_{1,p,q}^n \\ \Psi_{2,p+\frac{1}{2},q+\frac{1}{2}}^{n-\frac{1}{2}} \end{pmatrix} = 0, \text{ with} \quad (A2)$$

$$D_- = \frac{eA_y}{4\hbar}\left(\xi_q^{-1} + \xi_p^{-1} + \xi_p^{-1}\xi_q^{-1} + 1\right) +$$

$$\left(\frac{1}{2\Delta_x} - i\frac{1}{2\Delta_y}\right) - \left(\frac{1}{2\Delta_x} + i\frac{1}{2\Delta_y}\right)\xi_p^{-1} + \left(\frac{1}{2\Delta_x} + i\frac{1}{2\Delta_y}\right)\xi_q^{-1} - \left(\frac{1}{2\Delta_x} - i\frac{1}{2\Delta_y}\right)\xi_p^{-1}\xi_q^{-1}$$

(A3)

$$D_+ = -\frac{eA_y}{4\hbar}\left(\xi_q + \xi_p + \xi_p\xi_q + 1\right) +$$

$$\left(\frac{1}{2\Delta_x} + i\frac{1}{2\Delta_y}\right)\xi_p\xi_q - \left(\frac{1}{2\Delta_x} - i\frac{1}{2\Delta_y}\right)\xi_q + \left(\frac{1}{2\Delta_x} - i\frac{1}{2\Delta_y}\right)\xi_p - \left(\frac{1}{2\Delta_x} + i\frac{1}{2\Delta_y}\right)$$

. (A4)

To verify the stability of the algorithm we calculate the characteristic equation of the problem, obtained from the kernel of Eq. (A2), which is given by

$$\frac{1}{\lambda}\left(\lambda - 1 - \frac{V}{i\hbar 2}\Delta_t(\lambda+1)\right)^2 - (v_F \Delta_t)^2 D_- D_+ = 0, \quad (A5)$$

Using $\xi_p = e^{i\theta_p}$ and $\xi_q = e^{i\theta_q}$ it can be shown that:

$$D_- D_+ = \frac{1}{\Delta_x^2 \Delta_y^2}\left[-4\Delta_y^2 \cos^2\frac{\theta_q}{2}\sin^2\frac{\theta_p}{2} - \Delta_x^2 \cos^2\frac{\theta_p}{2}\left(\frac{eA_y}{\hbar}\Delta_y \cos\frac{\theta_q}{2} + 2\sin\frac{\theta_q}{2}\right)^2\right]. \quad (A6)$$

The nontrivial solutions $\lambda$ of characteristic equation are then:

$$\lambda = \frac{1}{(C-i)^2}\left[-1-C^2+\frac{B^2}{2}\pm B\sqrt{\left(\frac{B}{2}\right)^2-C^2-1}\right], \tag{A7}$$

with $C=\frac{V}{2\hbar}\Delta_t$ and $B^2=-(v_F\Delta_t)^2 D_- D_+ > 0$ real-valued parameters. From this result, it is simple to check that if $\left(\frac{B}{2}\right)^2-C^2-1<0$ then

$$|\lambda|=\frac{1}{(1+C^2)}\left[\left(1+C^2-\frac{B^2}{2}\right)^2+B^2\left(1+C^2-\frac{B^2}{4}\right)\right]^{1/2}=1. \tag{A8}$$

Thus, the algorithm is stable when $\left(\frac{B}{2}\right)^2-C^2-1<0$. This condition is equivalent to

$$\Delta_y^2\cos^2\frac{\theta_q}{2}\sin^2\frac{\theta_p}{2}+\Delta_x^2\cos^2\frac{\theta_p}{2}\left(\frac{eA}{2\hbar}\Delta_y\cos\frac{\theta_q}{2}+\sin\frac{\theta_q}{2}\right)^2 < \frac{\Delta_x^2\Delta_y^2}{(v_F\Delta_t)^2}\left(1+\left(\frac{V}{2\hbar}\Delta_t\right)^2\right) \tag{25}$$

The above inequality should be satisfied for all $\theta_p$ and $\theta_q$. In particular, it is enough to ensure that:

$$\frac{1}{\Delta_x^2\Delta_y^2}\left(\Delta_y^2+\Delta_x^2\left(\frac{eA}{2\hbar}\Delta_y+1\right)^2\right)(v_F\Delta_t)^2 < 1 \tag{26}$$

If we consider equally spaced nodes, i.e. $\Delta_x=\Delta_y=\Delta$ and time steps given by $\Delta_t=\frac{\Delta}{v_F}\alpha$, where $\alpha$ is a real valued positive constant, this condition is equivalent to:

$$\alpha < \sqrt{\frac{1}{1+\left(\frac{eA}{2\hbar}\Delta+1\right)^2}} \tag{27}$$

Hence, for these superlattices the numerical algorithm stability will depend on the amplitude of the magnetic vector potential. In case there is no magnetic vector potential

($A = 0$) we regain the usual formula for the stability of the FDTD algorithm $\Delta_t < \frac{\Delta}{v_F \sqrt{2}}$ [17, 18].

**Acknowledgements**

This work was partially funded by the Institution of Engineering and Technology (IET) under the A F Harvey Research Prize 2018, by the Simons Foundation, by Fundação para Ciência e a Tecnologia under Project PTDC/EEITEL/4543/2014 and by Instituto de Telecomunicações under Project No. UID/EEA/50008/2021. D. E. Fernandes acknowledges support by FCT, POCH, and the cofinancing of Fundo Social Europeu under the fellowship SFRH/BPD/116525/2016.